\documentclass[aps,prb,showpacs,amsmath,amssymb,floatfix,twocolumn]{revtex4}

\usepackage{graphicx}
\usepackage[percent]{overpic} % inset graphics
\usepackage{bm} % bold mathematical symbols
\usepackage{gensymb} % for degree

\begin{document}

\title{Study of pressure-induced amorphization in sulfur using \textit{ab initio} molecular dynamics}
\author{Du\v{s}an Pla\v{s}ienka}
\email{dusan.plasienka@fmph.uniba.sk}
\affiliation{Department of Experimental Physics, Comenius University,
\\ Mlynsk\'{a} Dolina F2, 842 48 Bratislava, Slovakia}

\author{Roman Marto\v{n}\'{a}k}
% \email{martonak@fmph.uniba.sk}
\affiliation{Department of Experimental Physics, Comenius University,
\\ Mlynsk\'{a} Dolina F2, 842 48 Bratislava, Slovakia}


% 07.05.Tp - Computer modeling and simulation
% 61.43.Bn - Molecular dynamics calculations in structural modeling of disordered solids
% 61.43.-j - Short-range order in amorphous materials
% 64.70.kg - solid-solid transitions

% 61.43.Dq - Amorphous semiconductors, metals, and alloys
% 71.15.Pd - Molecular dynamics calculations in electronic structure of solids
% 81.40.Vw - structural properties of materials

\date{\today}

\begin{abstract}
  We report results of ab initio constant-pressure molecular dynamics
  simulations of sulfur compression leading to structural transition and
  pressure-induced amorphization. Starting from the orthorhombic S-I phase
  composed of S$_8$ ring molecules we find at room temperature and pressure
  of 20 GPa a transformation to monoclinic phase where half of the
  molecules develop a different conformation. Upon further compression, the
  monoclinic phase undergoes pressure-induced amorphization into an
  amorphous phase, in agreement with experiments.  We study the dynamics of
  the amorphization transition and investigate the evolution of intra and
  intermolecular distances in the monoclinic phase in order to provide a
  microscopic insight into the rings disintegration process leading to
  amorphization.  In the amorphous form we examine the structural
  properties and discuss its relation to the experimentally found amorphous
  form and to underlying crystal phases as well.  The amorphous form we
  find appears to correspond to the experimentally observed low density
  amorphous form.
\end{abstract}

\maketitle

\section{Introduction}

Pressure is a key external variable determining structure and properties of
solids.  The most dramatic effect induced by pressure are structural
transformations between different crystalline phases. Providing access to
number of polymorphs, pressure-induced structural transitions are of
academic as well as practical interest in solid-state physics and materials
science. Besides transitions between stable or metastable crystalline
forms, transitions from crystal to amorphous form have been observed in
various materials, such as H$_2$O \cite{Mishima-1}, Si \cite{Deb-Si},
SiO$_2$ \cite{Hemley-SiO2}, etc.  The process has been called
pressure-induced amorphization (PIA) and since it was first studied in
compressed ice \cite{Mishima-1} its nature as well as the character of the
amorphous form created in this way has been extensively discussed (see
Refs.  \cite{Arora-PIA-review, Sharma-Sikka-PIA-review} for a detailed
review on this topic). Open questions include a number of issues. First is
the connection of the amorphous form created in this way to liquid or
crystalline phase. It was recognized that it might be structurally
connected either to higher-temperature liquid phase (and its corresponding
glass), or, alternatively, could represent a disordered version of some
underlying crystalline phase. Another discussion concerns the mechanism of
PIA and its thermodynamical description. PIA has been originally explained
as metastable melting, recognizing that compressed ice amorphized upon
crossing the negatively sloped melting line of water extrapolated to low
temperatures \cite{Mishima-1}. Another scenario referred as mechanical
melting is based on observation that the structure collapse might be driven
by elastic or lattice instabilities (by softening of certain elastic or
phonon shear modes) at high pressure conditions
\cite{Binggeli-Chelikowsky-SiO2, Brazhkin}. Upon approaching PIA, it has
been commonly observed that the x-ray diffraction patterns become less
crystalline suggesting that creation of defects often precede the
phenomenon of PIA.

In some materials the existence of more amorphous forms that differ in
density and microscopic structure has been observed. This phenomenon was
called polyamorphism, analogously to polymorphism. It has been found that
at least two different forms exist which have been called
low-density-amorphous (LDA) and high-density-amorphous (HDA) forms.
Polyamorphism was first observed in compressed ice \cite{Mishima-2} and
since then has been experimentally and theoretically studied in a number of
other common elements and compounds like Si \cite{Deb-Si, Daisenberger-Si},
Ge \cite{Principi-Ge}, SiO$_2$ \cite{Huang-Kieffer-SiO2-1, Trachenko-Dove},
etc.  While in some cases the amorphous-amorphous transition (AAT) was
found to be sharp \cite{Mishima-2, Daisenberger-Si, Principi-Ge}, in other
systems it was observed to proceed gradually \cite{Huang-Kieffer-SiO2-1,
  Durandurdu-GeSe2}. The sharpness of the transition might also be
temperature dependent, as observed, e.g. in SiO$_2$, where densification is
promoted at elevated temperatures in certain pressure region associated
with the ''reversibility window'' \cite{Trachenko-Dove, Trachenko}.
Similarly to the case of PIA, it has been found that there could be a
connection of AAT to liquid-liquid transition at higher temperatures or to
thermodynamical crystal-crystal transformation (see Refs.
\cite{HPC-Sanloup, McMillan-AAT-review-1, McMillan-AAT-review-2} for a
review on the phenomenon of AAT).

\begin{figure}
\includegraphics[width=\columnwidth]{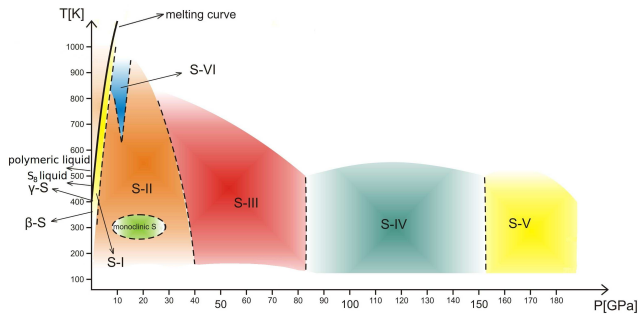}
\caption{(Color online) Thermodynamic phase diagram of sulfur. True
thermodynamic stability regions of some phases are uncertain because
large hysteresis on compression/decompression has been observed in the experiments
\cite{Deg-1,Deg-2}. Existence of some other crystalline phases is proposed
by numerical simulations \cite{Oganov, Pastorino-Gamba-1, Rudin-Liu}.}
\label{fig:S_phase_diagram}
\end{figure}

Recently the existence of PIA as well as of polyamorphism has been reported
also in sulfur \cite{Luo-Ruoff, Sanloup, Gregoryanz-unpublished}.  Sulfur
is one of the most common and important elements and its crystal structure
at ambient conditions belongs to the most complex ones found among pure
elements. Sulfur was experimentally studied in the pressure range from 0 to
230 GPa and at least ten different stable crystal structures have been
identified. Thermodynamic phase diagram of sulfur is presented in Fig.
\ref{fig:S_phase_diagram}. The diagram is based on data from number of
experiments \cite{Meyer, Deg-1, Deg-2, Deg-3, Deg-4, Deg-5, Hejny,
  Crapanzano, Zakharov-Cohen, Luo-Ruoff}.

The stable structure of sulfur at ambient conditions is orthorhombic
structure S-I which is highly complex as its unit cell consists of 128
atoms in 16 S$_8$ ring molecules. At temperatures close to melting curve,
S-I transforms to S-II upon pressure increase at 1.5 GPa and if thereafter
quenched to room temperature, S-II transforms further to S-III at 36 GPa
\cite{Deg-2}.  The structures of S-II and S-III are similar - both are
polymeric and consist of chains.  S-II is trigonal and formed by molecules
with shape of triangular chains and S-III is tetragonal with square-shaped
chains \cite{Fujihisa, Deg-1, Deg-2, Deg-5}.  Both structures are very
different from the molecular S-I phase and therefore one can expect that
they are separated from S-I by high energy barriers.  The complex
transition mechanism between molecular and polymeric sulfur needs to
include bond breaking in S$_8$ molecules and complete reorganization of the
entire structure.

At pressures over 83 GPa, S-III transforms into aperiodic incommensurately
(IC) modulated monoclinic phase S-IV \cite{Hejny, Deg-4, Nishikawa}.
Further increase of pressure above 135 GPa results into formation of phase
S-V with rhombohedral $\beta$-Po structure \cite{Deg-3, Deg-4, Deg-5,
  Luo-Greene-Ruoff}. The high-pressure phases of sulfur exhibit different
electrical properties. While low-pressure phases up to S-II are insulating,
S-III is a semiconductor and S-IV and higher-pressure phases are found to
be metallic and superconducting \cite{Nishikawa, Nishikawa-Niizeki-Shindo,
  Zakharov-Cohen, Luo-Vohra, Dunn-Bundy, Gregoryanz-supercond}.

Sulfur at ambient pressure melts at 115$\degree$C \cite{Meyer} and then
upon rising temperature undergoes a liquid-liquid transition at around
160$\degree$C \cite{Bellissent, Kalampounias, Scopigno} from S$_8$
molecular liquid to viscous polymeric liquid with rubberlike properties
\cite{Monaco} that is metallic \cite{Springborg}.  The process of
ring-opening polymerization resulting into creation of helical chains was
simulated in Refs.\cite{Tse-Klug-S, Jones-Ballone-2}. By rapid quenching of
this sulfur melt, amorphous version of solid sulfur that contains polymeric
chains is created.  Very stable version of amorphous sulfur prepared by a
rapid compression to 2 GPa has been recently reported \cite{Yu-S}.

Because of high barriers separating S-I from polymeric phases one can
expect strong kinetic effects and metastability, in particular at low
temperatures. Indeed, it was found that at room temperature compression of
S-I results in a transformation to S-III only when pressure of 36 GPa is
reached, thus completely skipping the S-II structure \cite{Deg-2}.
Metastability of the S-I molecular phase is found also in all experiments
observing PIA, upon approaching the pressure of amorphization
\cite{Luo-Ruoff, Sanloup, Gregoryanz-unpublished}.

Luo and Ruoff \cite{Luo-Ruoff} compressed S-I at room temperature and found
a transition to a monoclinic phase at about 5 GPa. It was, however, not
possible to determine the exact structure of this phase. Upon further
increase of pressure they observed a reversible amorphization starting at
18 GPa and completed at 25 GPa. Between 18 and 25 GPa they observed
significant decrease in intensity and increase in width of diffraction
peaks resulting in lower number of diffraction peaks observed.
Recrystallization to an unknown phase was observed at 37 GPa
\cite{Luo-Greene-Ruoff}. Similar results were observed in Ref. \cite{Akahama}.

In Ref. \cite{Luo-Ruoff} authors proposed two possible mechanisms of PIA.
In the first scenario the system attempts to transform into new structure
but remains trapped in disordered state before completing the transition
because of insufficient mobility of the atoms which does not allow the
reorganization of the structure. The second scenario represents
amorphization triggered by intramolecular bond breaking.

In Ref. \cite{Gregoryanz-unpublished}, Gregoryanz et al. studied
amorphization of sulfur at room temperature and below. They also observed
that diffraction reflections first start to broaden and decrease in
intensity at 25 GPa between 80 K and 175 K. Subsequently, PIA takes place
and is completed at 47 GPa at 80 K, at 45 GPa at 175 K and at 37 GPa at 300
K. In addition to PIA, Ref. \cite{Sanloup} reported the observation of LDA-HDA
polyamorphic transition in sulfur above 65 GPa at temperature 40 K.
According to the density and coordination number measured for these forms,
authors suggested that the LDA and HDA forms might correspond to their
crystalline counterparts, namely polymeric S-III and metallic S-IV. They
also pointed out that there is a crossing of the S-III/S-IV phase boundary
behind the LDA-HDA transition. However, they also admit the possibility
that besides a genuine polyamorphism the experimental data might
be also compatible with the creation of small nanocrystals in the sample.
This cannot be distinguished from the genuine amorphous form within the
resolution of the x-ray diffraction technique.

In this paper we aim at resolving the open questions concerning the PIA and
polyamorphism in sulfur by means of ab initio constant-pressure molecular
dynamics. Both Refs. \cite{Luo-Ruoff, Gregoryanz-unpublished} indicate that
before PIA the structure undergoes substantial changes. It is plausible to
assume that in the conditions of strong overpressurization the S$_8$
molecules do not remain intact and become distorted even before the onset
of PIA. Since it was not possible to determine the precise character of
these structural changes experimentally, it appears useful to complement
the experiments by computer simulation. This could also shed light on the
microscopic mechanism of the PIA and help to understand the subsequent
polyamorphic transition. The paper is organized as follows. In section 2 we
present the simulation method. In section 3 we describe the simulation
protocol and discuss the results and compare them to experimental data. In
the final section we draw some conclusions.

\section{Simulation methods and results}

To perform ab initio molecular dynamics (MD) simulations we used VASP package
\cite{VASP-1, VASP-2, VASP-3, VASP-4}. To simulate system under
constant pressure, we used the idea based on the Berendsen barostat
\cite{Berendsen}. After performing 20 MD steps with time step of 2 fs in
constant supercell (total simulated time of 40 fs), we rescaled parameters
of the supercell according to the difference of external pressure
$\textbf{P}_{ext}$ and instantaneous internal pressure in the system
$\textbf{P}_{int}$ following the Berendsen scheme.  We note that our
procedure is slightly different from the original one \cite{Berendsen}
because we do not apply the scaling at every MD step.

The cell matrix $\mathbf{h}=(\vec{a},\vec{b},\vec{c})$ where the three
vectors $\vec{a},\vec{b},\vec{c}$ span the simulation supercell, together
with atomic positions $\textbf{r}_i$ and velocities $\textbf{v}_i$ were
transformed by the scaling matrix \bm{$\mu$} following the rule

\begin{center}
$\textbf{h} \rightarrow$ \bm{$\mu$}$\textbf{h}$, $\textbf{r}_i
\rightarrow$ \bm{$\mu$}$\textbf{r}_i$, $\textbf{v}_i \rightarrow $
\bm{$\mu$}$\textbf{v}_i$,

\bm{$\mu$} $ = \textbf{1} - \frac{\beta \Delta t}{3\tau_P}
\left(\textbf{P}_{ext} - \textbf{P}_{int}\right) $,
\end{center}

where $\beta$ is the bulk modulus of the system, $\Delta t$ is the time
step of the transformation (40 fs in our case) and $\tau_P$ is the
relaxation time scale defining how quickly the algorithm responds to
pressure fluctuations (only the $\frac{\beta \Delta t}{3\tau_P}$ ratio is
relevant).

All simulations were performed on a sample consisting of 512 atoms in a
supercell (generated as $2 \times 2 \times 1$ supercell of S-I unit cell)
with periodic boundary conditions.  The core electrons were dealt with the
projector augmented wave (PAW) pseudopotential method \cite{PAW, VASP-PAW}.
Each atom contributed to the electronic problem with 6 electrons from
$3s^2$ and $3p^4$ sulfur valence orbitals. These electrons were treated by
means of density functional theory (DFT) within the generalized gradient
approximation (GGA) in the Perdew-Burke-Ernzerhof \cite{PBE} (PBE) scheme.
The Kohn-Sham equations \cite{KS} were solved in a plane wave basis set
with energy cutoff of 360 eV. Since the supercell was fairly large ($22
\times 27 \times 25$ \AA \,at 0 GPa) the k-point grid was well approximated
by taking only the $\Gamma$-point.

\section{Simulation results and discussion}

\subsection{Simulation protocol}
% \label{sec:simulation-protocol}

\begin{figure}
\includegraphics[width=\columnwidth]{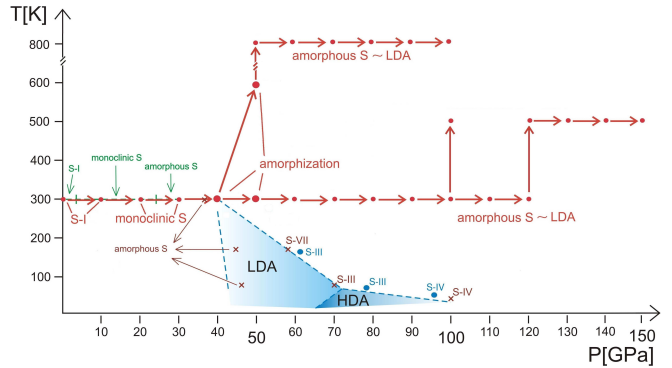}
\caption{(Color online) Simulation protocol of the present study (red dots
  and arrows) along experimental data for amorphous sulfur from Ref.
  \cite{Luo-Ruoff} (green), Ref. \cite{Gregoryanz-unpublished} (brown)
  and Refs. \cite{Sanloup, HPC-Sanloup} (LDA and HDA forms in light and
  dark blue).  Three amorphization points are marked by bold circles.}
\label{fig:Simulation_protocol}
\end{figure}

The simulation protocol together with the experimental data for amorphous
sulfur from Refs. \cite{Luo-Ruoff, Sanloup, Gregoryanz-unpublished} is
schematically shown on Fig. \ref{fig:Simulation_protocol}. We started the
simulation from the optimized S-I structure at 0 GPa and 0 K and initially
heated the system at zero pressure to 300 K.  Afterwards we gradually
increased pressure in 10 GPa steps keeping the temperature at 300 K. We
allowed the system to equilibrate for 4 ps at pressures up to 30 GPa and
for 10-30 ps at higher pressures where substantial structural changes take
place. At 20 GPa we observed a transition to a new molecular phase with
monoclinic lattice formed by S$_8$ molecules with two different
conformations.  Upon further increase of pressure to 40 GPa we observed
initial creation of structural defects in molecular structure where few
bond interchanges between nearby molecules were present. From this point, we
proceeded with simulation along three different paths. We extended the 40
GPa simulation up to nearly 30 ps which gave us access to detailed
information about early stages of PIA. Second, after 4 ps run at 40 GPa, we
increased pressure to 50 GPa and beyond, keeping the system at room
temperature. Third, we increased both pressure to 50 GPa and temperature to
600 K in order to further accelerate PIA. PIA in our simulations was
therefore observed at three different P-T conditions. The amorphous state
created at 50 GPa thereafter persisted to very high P-T points (150 GPa,
500 K) and (100 GPa, 800 K) with no sign of progress in recrystallization
or transition to different amorphous form (Fig.
\ref{fig:Simulation_protocol}).

In the following subsections we discuss in detail the properties of the
monoclinic phase, analyze the origin and dynamics of the PIA transformation
and study the properties of the amorphous form. We shall denote the monoclinic
and amorphous forms found in our simulations as m-S and a-S.

\subsection{Monoclinic sulfur}

As we increased pressure from 10 GPa to 20 GPa, the supercell of S-I
distorted by lowering the $\alpha$ angle from 90$\degree$ to 86.5$\degree$
at 40 GPa according to MD simulations at 300 K and subsequent geometric
optimization at zero temperature. This change of supercell is compatible
with change of lattice symmetry from orthorhombic to monoclinic. At the
same time half of the originally identical S$_8$ molecules
deformed to less symmetric form. We denote these deformed S$_8$ molecules
in m-S as type B and the original ones as type A.

\begin{figure}
\includegraphics[width=\columnwidth]{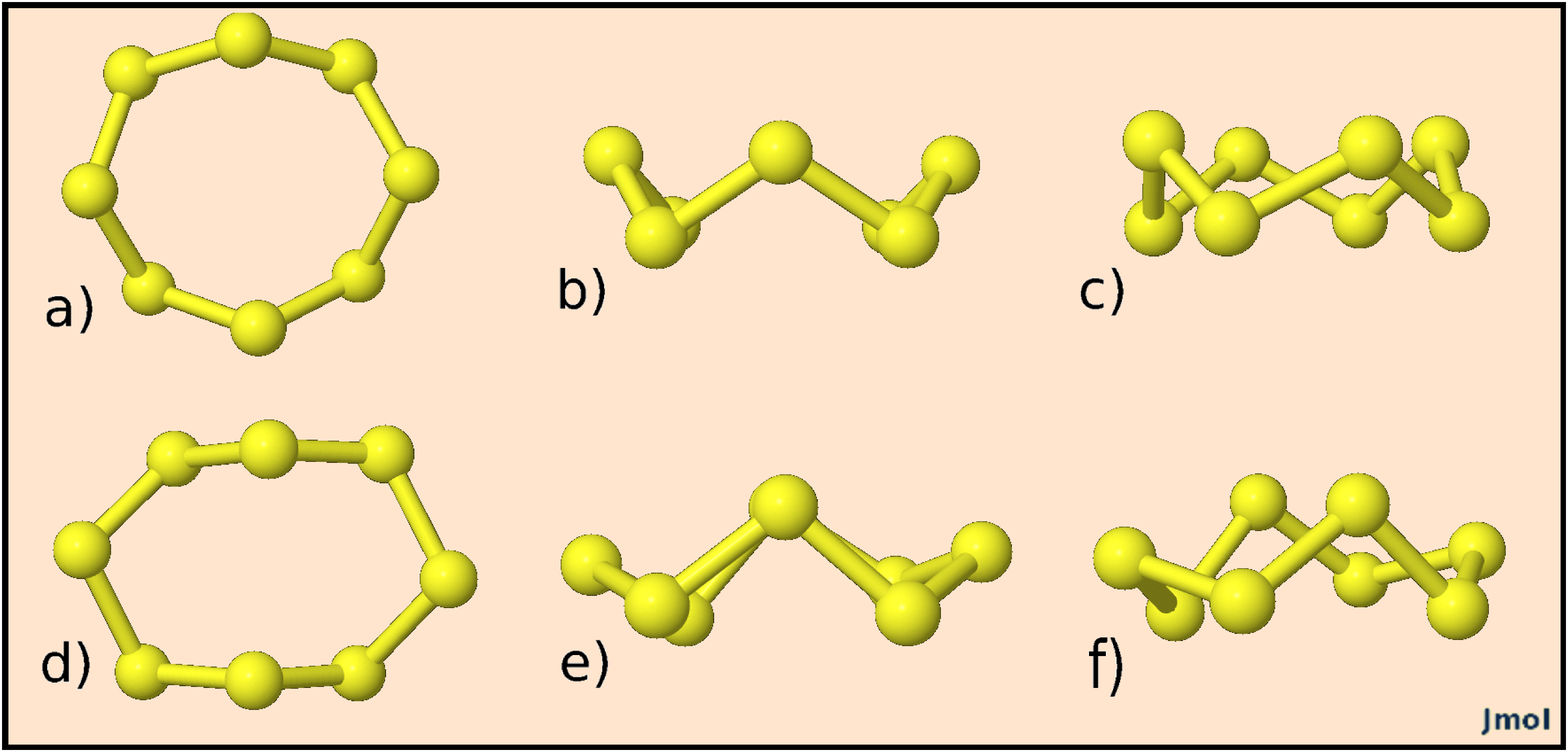}
\includegraphics[width=\columnwidth]{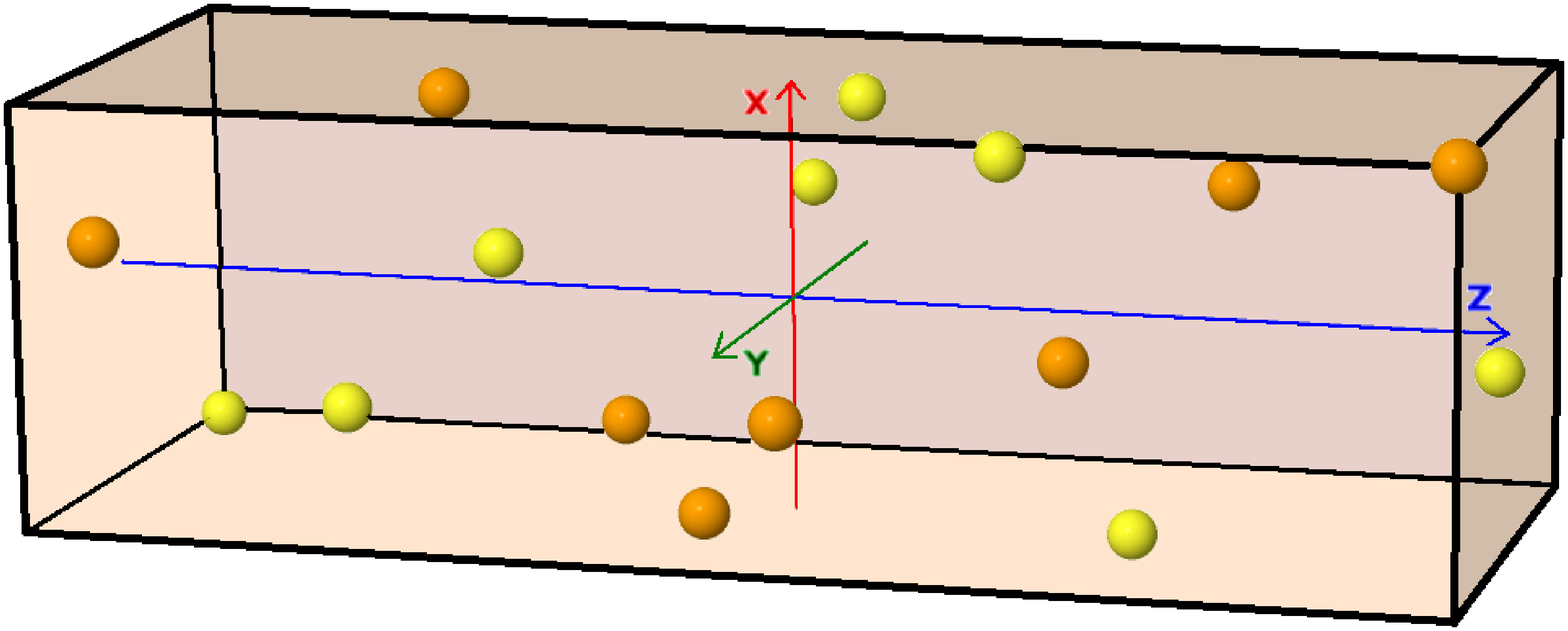}
\caption{(Color online) Comparison of type A - a)-c) and type B - d)-e)
  molecules and visualization of m-S structure. Yellow (lighter) spheres
  represent mass centers of original type A molecules and orange (darker)
  spheres represent centers of deformed type B molecules.  Pictures were
  generated by Jmol \cite{Jmol}.}
\label{fig:Molecules}
\end{figure}

Type A molecules are naturally most common 8-atomic ring-puckered $D_{4d}$
isomer with crown shape \cite{Meyer, Jones-Ballone-1, Hohl-Jones}. This
isomer forms $\alpha$, $\beta$ and $\gamma$-S structures and is also
present in liquid sulfur.  Type B molecules originated from deformation of
type A molecules and possess lower $C_2$ symmetry. Their shape is shown and
compared to the $D_{4d}$ isomer in Fig. \ref{fig:Molecules} together with
the schematic view on the structure of m-S.

As far as m-S appears to be a new structure of sulfur, we optimized the 128
atom cell representing quarter of the simulation supercell.
The relation of m-S to monoclinic-S found by Luo and Ruoff could not be
determined because structural data for this phase are not available. We were
not able to find the exact space group symmetry for this complex monoclinic
phase with large unit cell.

We also calculated the equation of states (EOS) of S-I and m-S to compare
their relative stability.  According to the calculated EOS, we found that
m-S at zero temperature is more stable than S-I at pressures greater than
29 GPa. We also found that optimization of the m-S unit cell at 50 GPa and
higher pressures resulted into formation of amorphous form directly during
structural optimization. This leads to the conclusion that 50 GPa is the
upper limit of the metastability of molecular m-S structure and beyond
50 GPa m-S cannot exist anymore.

\subsection{Mechanism and dynamics of pressure-induced amorphization}

\begin{figure}
\begin{overpic}[width=\columnwidth]{Volume.eps}
\put(43,12){\includegraphics[width=0.48\columnwidth]{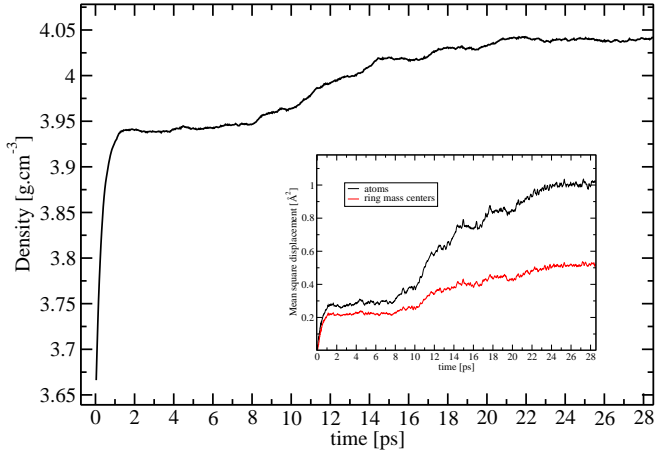}}
\end{overpic}
\caption{(Color online) Density relaxation during m-S$\rightarrow$a-S
  transition at 40 GPa.  Elastic compression of m-S completed in 1.5 ps is
  followed by 6.5 ps run when m-S remains crystalline. At 8 ps the
  amorphization starts and in the following 13 ps the system undergoes
  non-elastic volume compression. (Inset) Mean square displacement $\Delta
  r^2(t)$ as a function of time for atoms (upper black curve) and ring mass
  centers (lower red curve) for the entire MD run at 40 GPa. The MSD
  curves have clearly distinct character in the crystalline m-S, where MSD
  stabilizes at constant value until amorphization begins at 8 ps, and
  after 8 ps, where it starts to grow rapidly. The curves of the MSD still
  grow till the end of the run which means that the amorphization process
  is not yet completed and further density increase could be expected if
  considerably longer simulation times were available.}
\label{fig:Volumes}
\end{figure}

\begin{figure}
\includegraphics[width=\columnwidth]{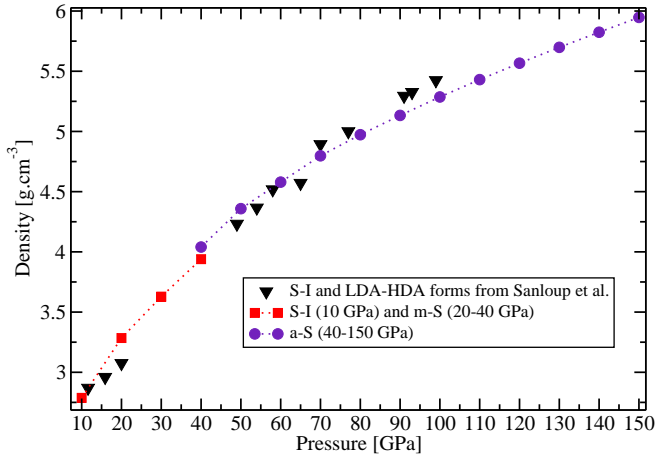}
\caption{(Color online) Density of sulfur under pressure according to the
  presented MD simulations (red squares for m-S and violet circles for a-S)
  and S-I and LDA-HDA densities from Refs. \cite{Sanloup, HPC-Sanloup}
  (black triangles). Our results are for 300 K for pressures 10-120 GPa and
  for 500 K for 130-150 GPa. The transition from m-S to a-S at 40 GPa is
  seen as a small jump in density.}
\label{fig:Density}
\end{figure}

The 30 ps simulation run at 40 GPa gives us access to detailed information
about the early stages of PIA.  On Fig. \ref{fig:Volumes} we show the
evolution of the system density from the beginning of compression (at 0 ps)
from 30 to 40 GPa till the end of the 40 GPa run. In this figure, the
elastic and non-elastic parts of the density increase can be clearly
recognized.  The increase of density from 3.63 to 3.94 g cm$^{-3}$ at 1.5
ps corresponds to the elastic compression of molecular m-S. At 1.5 ps, the
pressure in the system is equilibrated to external 40 GPa, and the m-S
structure in next 6.5 ps does not undergo any change.  At 8 ps, however,
the system spontaneously starts to amorphize and one can see a step-wise
decrease of volume persisting until 21 ps of the MD run. Volume reduction
during amorphization is accompanied by further decrease of the monoclinic
angle from 86.5$\degree$ to 83.2$\degree$ in a-S at the end of the MD run.
The resulting amorphous version of sulfur at 40 GPa has density of 4.04 g
cm$^{-3}$ that corresponds to density increase of 2.54\% from m-S at 40
GPa. It is plausible to assume that the compression would further continue
if longer simulation times were accessible. Amorphous versions of sulfur
obtained at pressure of 50 GPa and temperatures of 300 and 600 K were
investigated to much higher pressures (to 150 GPa in lower and to 100 GPa
in higher temperature branch). We did not observe any significant jump in
density that could be associated with the LDA to HDA polyamorphic
transition.  The graph of densities of sulfur from 10 to 150 GPa together
with the data from Refs.  \cite{Sanloup, HPC-Sanloup} is shown in Fig.
\ref{fig:Density}. Together with the increase of density, the process of
amorphization between 8 ps and 21 ps leads into an increase of average
energy while enthalpy decreases by 30 meV per particle.

Since during the amorphization transformation the atoms are likely to
diffuse over finite distances away from their original positions in the
crystalline phase we computed the time-dependent mean square displacement
$\Delta r^2(t)$ (MSD) of atoms and molecular mass-centers in order to
monitor the diffusion during amorphization. The values of $\Delta r^2(t)$
are evaluated as

\begin{center}
$\Delta r^2(t) = \left< \left(\bm{r}(t) - \bm{r}_0 \right)^2 \right>_N$,
\end{center}

where $\bm{r}_{0,i}$ are the initial positions which are subtracted from
the actual ones at every time step. We have plotted the time-dependent MSD
of atoms and mass-centers for m-S and a-S at 40 GPa in the inset of Fig.
\ref{fig:Volumes}.  From the graph, one can see a clear difference between
the character of MSD in crystalline and amorphous sulfur in the time
interval where both phases exist. The crystalline case is recognized by a
constant value of MSD between 1.5 and 8 ps, while during the amorphization,
MSD of atoms and mass-centers grows even after 28 ps. This shows that
despite volume is stabilized at 21 ps, the structure continues to evolve as
amorphization proceeds. This also points to the intrinsic time-scale
limitation of the ab initio study since following the evolution of this
fairly large system over substantially longer time, although very
desirable, would be prohibitively expensive.

In molecular crystals at low pressure the intermolecular distances between
atoms are typically much larger than the corresponding intramolecular ones.
Upon compression the former ones decrease as the molecules approach each
other and when the two kinds of distances become comparable a transition
from molecular to non-molecular, or polymeric phase, may take place.
Examples are N$_2$ \cite{Goncharov-N}, CO$_2$ \cite{Santoro-1}, etc. It
is plausible to assume that a similar scenario may apply here. In
particular, the PIA of the strongly overpressurized ring-molecular phase
could be triggered as the molecules approach each other closely.

In order to check this possibility we focused on the evolution of
intramolecular bond lengths and intermolecular distances upon increasing
pressure. Compression of m-S leads to considerable decrease in
intermolecular space, while the intramolecular bond lengths remain
practically unchanged. We also observed a lowering of the bond angles in
molecules from 107$\degree$ at 10 GPa to 97$\degree$ at 40 GPa.

\begin{figure}
\includegraphics[width=\columnwidth]{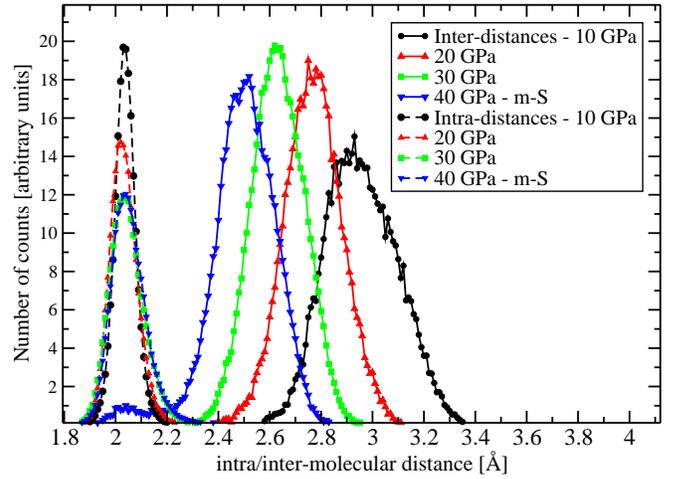}
\caption{(Color online) Comparison of bond length and nearest
  intermolecular distance distributions in S-I at 10 GPa and in m-S at
  20-40 GPa. The structure of m-S becomes unstable at 40 GPa where the distributions start to overlap. The structure of m-S at 40 GPa contains
  some bonds between different molecules (defects) that are represented by
  the small peak around the sulfur covalent diameter of 2.04 \AA. The
  distribution at 40 GPa is averaged over 6.5 ps interval of m-S
  existence.}
\label{fig:Inter-intradistances}
\end{figure}

In Fig. \ref{fig:Inter-intradistances} we present the distributions of the
nearest neighbors (n.n.) intramolecular distances (bond lengths) and the
nearest intermolecular distances in S-I and m-S at pressures from 10 to 40
GPa. Every molecule contributes with 8 values to both histograms and the
nearest intermolecular distance is defined as the closest distance between
the atom and all atoms in other molecules.

We see that up to the pressure of 30 GPa the two distributions are clearly
separated and do not overlap. Under these conditions the molecular phase
persists. At 40 GPa, we see that the intermolecular distribution develops a
small peak located around the sulfur bond length. This points to the
existence of structural defects in m-S where certain atoms from different
molecules start bonding.

Even without existence of defects, the figure reveals that at 40 GPa
molecules interact strongly as their intra and intermolecular distances
distributions start to overlap. We note that in experiment at room
temperature the PIA was observed at 37 GPa \cite{Gregoryanz-unpublished}
which is in perfect agreement with our results.  The data suggest
that PIA is likely to be primarily driven by the overlap of these two
distributions, similarly to other cases of molecular to non-molecular
transformation in crystals.

\begin{figure}
\includegraphics[width=\columnwidth]{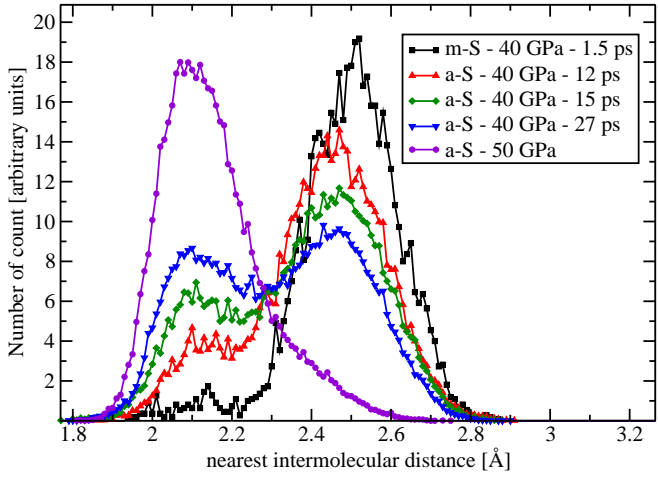}
\caption{(Color online) Evolution of the nearest intermolecular distance
  distribution during amorphization at 40 GPa. Starting from the defective
  m-S at 1.5 ps, the histogram moves left as more and more atoms make their
  bonds with surrounding molecules. Distribution at 50 GPa, where most of
  the atoms bond to other molecules, is shown for comparison.}
\label{fig:Interdistances_amorphization}
\end{figure}

This scenario is further confirmed by analyzing the dynamics of the early
stages of the PIA. We show in Fig. \ref{fig:Interdistances_amorphization}
the evolution of the nearest intermolecular distances histograms averaged
over short time intervals at 12, 15 and at 27 ps of the 40 GPa run. As
presented in the figure, the character of the nearest intermolecular
distance distribution considerably changes at 12 ps when it develops a
major peak weighted around the sulfur bond length. At the end of the 40 GPa
run, many of the atoms are already forming covalent bonds with atoms from
different - previously separated molecules.  By visual inspection we also
find that the amorphization process proceeds mainly around the original
structural defects.

\begin{figure}
\includegraphics[width=\columnwidth]{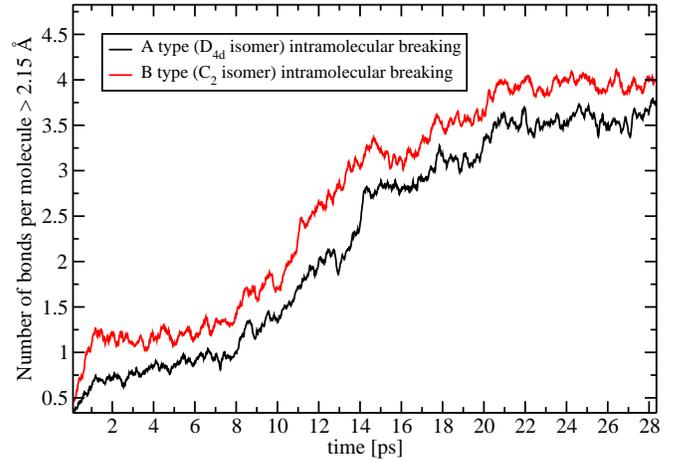}
\caption{(Color online) Number of intramolecular distances longer than 2.15
  \AA \,in type A molecules (black lower curve) and in type B molecules
  (red upper curve).  The numbers are normalized such that they represent
  an average number of bonds longer than the limit per one molecule of
  certain type. Amorphization at 40 GPa starting at 8 ps is represented by
  rapid growth of both curves as intramolecular bonds are progressively
  broken. At the end of the run, roughly half of the bonds in molecules is
  broken.}
\label{fig:Intra}
\end{figure}

In order to further clarify the amorphization transformation, we also study
the interactions between type A and type B molecules separately. In Fig.
\ref{fig:Intra} we show the evolution of the number of intramolecular
distances longer than 2.15 \AA \, for A and B molecules. (This limit has
been conventionally chosen and is 5.4 \% longer than sulfur covalent
diameter 2.04 \AA).  The number of bond lengths longer than 2.15 \AA \,in
type B molecules is always somewhat greater than in A molecules indicating
that type B molecules are more likely to develop bond breakings than A
molecules.

\begin{figure}
\includegraphics[width=\columnwidth]{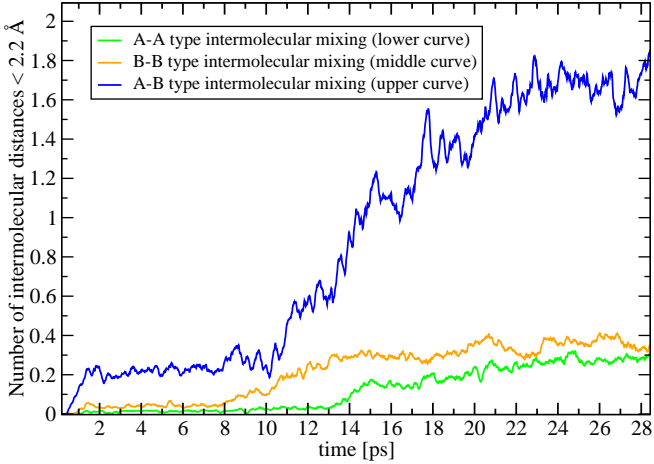}
\caption{(Color online) Number of intermolecular distances shorter than 2.2
  \AA \,between type A molecules (lower green curve), between type B
  molecules (middle red curve) and between A-B molecules mixed (blue upper
  curve). Amorphization starting at 8 ps corresponds to the first increase
  of the B-B curve followed by rapid A-B mixing starting at 10 ps and
  mixing between A molecules that starts another 3 ps later. The
  normalization is chosen such that the sum of all three graphs represents
  an average number of the intermolecular distances shorter than 2.2 \AA
  \,per one molecule. (Note that there are more than twice as many possible
  connections for A-B pairs than for A-A and B-B pairs).}
\label{fig:Inter}
\end{figure}

As a complementary information to bond lengths evolution, on Fig.
\ref{fig:Inter} we investigate the number of intermolecular distances
shorter than 2.2 \AA \,(close intermolecular approachings) for A-A, B-B and
A-B pairs separately. The figure confirms that B molecules are indeed more
involved in the early stages of amorphization which starts by a sudden
increase of the number of the B-B approachings at 8 ps.  Only after next
two picoseconds, molecules A and B start to mix together as the A-B curve
starts to grow after 10 ps. The mixing of A molecules starts even later, 3
ps after the beginning of A-B mixing.  At the end of the run at 40 GPa,
approximately 30 \% of all atoms have one atom from a different molecule
closer than 2.2 \AA, in average. This clearly shows that the amorphization
at 40 GPa proceeds slowly and even after 30 ps we still observe early
stages of the process.

Altogether, the analysis of bond lengths and intermolecular distances
evolution during amorphization provides interesting information about the
ring disintegration process leading to amorphization. In particular, we
identify the different role played by the A and B molecules.

In order to further characterize the transition from crystalline to
disordered structure, we also note that the mean intramolecular distance
between previously identified n.n. grew from 2.06 \AA \,in m-S at 40 GPa to
2.60 \AA \,in a-S at 50 GPa. Although this quantity no longer represents
any kind of bond length, it provides information about the amount of diffusion
in the system, in addition to the previously shown time-dependent MSD.

\subsection{Amorphous form}

S-I at ambient pressure is a soft material with $\beta$ = 7.7 GPa. This
reflects the presence of fairly large intermolecular space. The crystal to
amorphous form transition at 40 GPa is accompanied by sharp increase of
density (Figs.~\ref{fig:Volumes} and \ref{fig:Density}) possibly indicating
the first-order nature of the transition as suggested in Ref.
\cite{Sanloup}.

After obtaining a-S at 50 GPa, we performed further simulations at higher
pressures and also at higher temperatures (Fig.
\ref{fig:Simulation_protocol}). Even at the highest pressure of 150 GPa and
elevated temperature of 800 K at 100 GPa we have not found any evidence of
recrystallization or transition to a distinct amorphous form. On the
contrary, we have found a-S created at 50 GPa to remain without any serious
change of structure, except for slow equilibration of the first peak of the
radial distribution function (RDF) of a-S, as will be discussed later. The
lack of observation of recrystallization in our simulations was most likely
caused by short time scale of our simulations. In principle, the
non-observation of transition to HDA form, if this indeed exists, could be
related to the same time scale problem.

\begin{figure}
\includegraphics[width=\columnwidth]{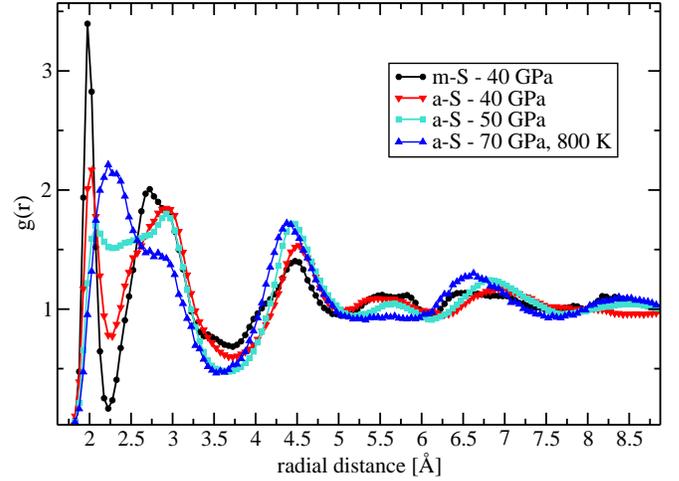}
\caption{(Color online) Comparison of RDFs of m-S at 40 GPa (black circled curve)
 and a-S at 40 GPa (red down triangles), at 50 GPa (turquoise
  squares) and at 70 GPa and 800 K (blue up triangles). Similarity of RDFs
  at 40 GPa reflects close relation of 40 GPa a-S to m-S. RDF of a-S at 70
  GPa corresponds quite well to the experimental g(r) in Ref.
  \cite{Sanloup} even though simulated g(r) is not fully equilibrated.}
\label{fig:RDF}
\end{figure}

In Fig. \ref{fig:RDF}, we present the RDFs of m-S at 40 GPa and a-S at 40,
50 and 70 GPa.  The decrease of the first peak maximum and the filling of
the first minimum is a consequence of breaking of intramolecular bonds. As
molecules start to disintegrate and make more bonds with other molecules,
the first two peaks start merging, although second peak remains
recognizable even to highest P-T conditions investigated in our
simulations. This persisting first peak separation indicates that some
short-range order of m-S remains present during simulated PIA, since there
is still some abundance of the next-to-n.n. distances. This implies that
our a-S still contains some fragments of the original S$_8$ molecules.

Next, we analyze the properties of a-S and its correspondence to the LDA
form from Ref. \cite{Sanloup}. We find that a-S and LDA are indeed very
similar, but the possible relation of a-S to some underlying crystalline
phase could not be clearly identified.

First, we observe that the RDF of a-S corresponds quite well to the RDF of
LDA from Ref. \cite{Sanloup}. We refer to the resemblance between the RDF
of LDA at 65 GPa (Ref. \cite{Sanloup}) and the RDF of a-S at 70 GPa shown
in blue in Fig. \ref{fig:RDF}. The agreement of the densities between a-S
and LDA is also very good, especially at 60 GPa.

Next, we find that the value of the coordination number in a-S at 50 GPa is
N$_C$ = 16.2 (for $r_c = 3.65$ \AA) and N$_C$ = 16.1 at 70 GPa ($r_c = 3.5$
\AA) which is again in close agreement with the experimental value of 16.1
at 65 GPa from Ref. \cite{Sanloup}. This suggests that our a-S is indeed
similar to the experimentally observed LDA form.

Now, we discuss the suggestion put forward in Ref. \cite{Sanloup}, namely
that the LDA form might be structurally related to crystal phase S-III.
This suggestion was based on the comparison of density and coordination
number, which is 17 in S-III for the same $r_c$ radius as for LDA. This
relation also seems plausible taking into account the positions of the two
forms in the phase diagram.

While N$_C$ in the S-III phase calculated to the first LDA minimum is 17,
we found that N$_C$ for m-S at 40 GPa with cutoff $r_c$ = 3.8 \AA \,(which
is the first minimum of a-S at 40 GPa), equals 15.9. We note that in case
of m-S the distance of 3.8 \AA \,naturally corresponds to the second RDF
minimum at 40 GPa.  Therefore the simulated a-S could also be structurally
related to its parent phase m-S, rather than to S-III.

Change of the structure of sulfur under pressure is also accompanied by
change of its electronic properties. We found that the amorphization at 40
and at 50 GPa is also accompanied by metallization. According to the
computed electronic density of states ($e$DOS), we found m-S to be an
insulator up to 30 GPa with energy band gap of 0.4 eV at 20 GPa and 0.1 eV
at 30 GPa. At 40 GPa, we found a small elevation of the $e$DOS minimum at
Fermi energy to a non-zero value before amorphization of m-S started. In
case of a-S at 40 GPa and at higher pressures, we find the amorphous form
to be metallic. Due to the well-known problem of underestimation of energy
gap in DFT calculations using approximate exchange-correlation functionals
it is possible that the true gaps in respective phases might be larger.

\section{Conclusions}

We performed ab initio constant-pressure MD simulations of elemental sulfur
on a sample containing 512 atoms. We observed two transitions - from S-I to
m-S at 20 GPa and from m-S to a-S at 40 and 50 GPa. The structure of m-S is
similar to S-I and might correspond to monoclinic-S observed by Luo and
Ruoff in Ref. \cite{Luo-Ruoff}. It consists of distorted (called type B)
and undistorted (type A) S$_8$ molecules.  While the a-S form we found
appears rather similar to the experimentally found LDA form from Ref.
\cite{Sanloup}, we did not find a subsequent transition to HDA form. This
could either reflect a too short time scale of our simulations, or the
possibility put forward in Ref.\cite{Sanloup} that the observed LDA and HDA
form might actually have nanocrystalline structure. The possible relation
of a-S to underlying crystal phase is not clear since we found similarities
to both m-S and S-III. The density-driven amorphization process starts at
40 GPa when the distributions of nearest intra and intermolecular distances
begin to overlap. This leads to bond interchanges and eventually to
molecular disintegration and formation of structurally disordered phase. We
found that in the early stages of the amorphization process the B molecules
are substantially more involved than A molecules.  It would be interesting
to obtain, if possible, high-quality diffraction pattern for the
crystalline structure before PIA and compare it to our m-S.

\begin{acknowledgments}
This work was supported by the Slovak Research and Development Agency
under Contract No. APVV-0558-10 and by the project implementation
26220220004 within the Research \& Development Operational Programme
funded by the ERDF.
\end{acknowledgments}

% \bibliographystyle{apsrev}
% \bibliography{references}

\begin{thebibliography}{63}
\expandafter\ifx\csname natexlab\endcsname\relax\def\natexlab#1{#1}\fi
\expandafter\ifx\csname bibnamefont\endcsname\relax
  \def\bibnamefont#1{#1}\fi
\expandafter\ifx\csname bibfnamefont\endcsname\relax
  \def\bibfnamefont#1{#1}\fi
\expandafter\ifx\csname citenamefont\endcsname\relax
  \def\citenamefont#1{#1}\fi
\expandafter\ifx\csname url\endcsname\relax
  \def\url#1{\texttt{#1}}\fi
\expandafter\ifx\csname urlprefix\endcsname\relax\def\urlprefix{URL }\fi
\providecommand{\bibinfo}[2]{#2}
\providecommand{\eprint}[2][]{\url{#2}}

\bibitem[{\citenamefont{Mishima et~al.}(1984)\citenamefont{Mishima, Calvert,
  and Whalley}}]{Mishima-1}
\bibinfo{author}{\bibfnamefont{O.}~\bibnamefont{Mishima}},
  \bibinfo{author}{\bibfnamefont{L.~D.} \bibnamefont{Calvert}},
  \bibnamefont{and} \bibinfo{author}{\bibfnamefont{E.}~\bibnamefont{Whalley}},
  \bibinfo{journal}{Nature} \textbf{\bibinfo{volume}{310}},
  \bibinfo{pages}{939} (\bibinfo{year}{1984}).

\bibitem[{\citenamefont{Deb et~al.}(2001)\citenamefont{Deb, Wilding,
  Somayazulu, and McMillan}}]{Deb-Si}
\bibinfo{author}{\bibfnamefont{S.~K.} \bibnamefont{Deb}},
  \bibinfo{author}{\bibfnamefont{M.}~\bibnamefont{Wilding}},
  \bibinfo{author}{\bibfnamefont{M.}~\bibnamefont{Somayazulu}},
  \bibnamefont{and} \bibinfo{author}{\bibfnamefont{P.~F.}
  \bibnamefont{McMillan}}, \bibinfo{journal}{Nature}
  \textbf{\bibinfo{volume}{414}}, \bibinfo{pages}{528} (\bibinfo{year}{2001}).

\bibitem[{\citenamefont{Hemley et~al.}(1988)\citenamefont{Hemley, Jephcoat,
  Mao, Ming, and Manghnani}}]{Hemley-SiO2}
\bibinfo{author}{\bibfnamefont{R.~J.} \bibnamefont{Hemley}},
  \bibinfo{author}{\bibfnamefont{A.~P.} \bibnamefont{Jephcoat}},
  \bibinfo{author}{\bibfnamefont{H.~K.} \bibnamefont{Mao}},
  \bibinfo{author}{\bibfnamefont{L.~C.} \bibnamefont{Ming}}, \bibnamefont{and}
  \bibinfo{author}{\bibfnamefont{M.~H.} \bibnamefont{Manghnani}},
  \bibinfo{journal}{Nature} \textbf{\bibinfo{volume}{334}}, \bibinfo{pages}{52}
  (\bibinfo{year}{1988}).

\bibitem[{\citenamefont{Arora}(2002)}]{Arora-PIA-review}
\bibinfo{author}{\bibfnamefont{A.~K.} \bibnamefont{Arora}}, in
  \emph{\bibinfo{booktitle}{High Pressure Phenomena}}, edited by
  \bibinfo{editor}{\bibfnamefont{R.~J.} \bibnamefont{Hemley}} \bibnamefont{and}
  \bibinfo{editor}{\bibfnamefont{G.~L.} \bibnamefont{Chiarotti}}
  (\bibinfo{publisher}{IOS PRESS}, \bibinfo{address}{Amsterdam, The
  Netherlands}, \bibinfo{year}{2002}), pp. \bibinfo{pages}{545--560}, ISBN
  \bibinfo{isbn}{1-58603-269-0}.

\bibitem[{\citenamefont{Sharma and Sikka}(1996)}]{Sharma-Sikka-PIA-review}
\bibinfo{author}{\bibfnamefont{S.~M.} \bibnamefont{Sharma}} \bibnamefont{and}
  \bibinfo{author}{\bibfnamefont{S.~K.} \bibnamefont{Sikka}},
  \bibinfo{journal}{Prog. Mat. Sci.} \textbf{\bibinfo{volume}{40}},
  \bibinfo{pages}{1} (\bibinfo{year}{1996}).

\bibitem[{\citenamefont{Binggeli and
  Chelikowsky}(1992)}]{Binggeli-Chelikowsky-SiO2}
\bibinfo{author}{\bibfnamefont{N.}~\bibnamefont{Binggeli}} \bibnamefont{and}
  \bibinfo{author}{\bibfnamefont{J.~R.} \bibnamefont{Chelikowsky}},
  \bibinfo{journal}{Phys. Rev. Lett.} \textbf{\bibinfo{volume}{69}},
  \bibinfo{pages}{2220} (\bibinfo{year}{1992}).

\bibitem[{\citenamefont{Brazhkin et~al.}(1997)\citenamefont{Brazhkin, Lyapin,
  Stalgorova, Gromnitskaya, Popova, and Tsiok}}]{Brazhkin}
\bibinfo{author}{\bibfnamefont{V.~V.} \bibnamefont{Brazhkin}},
  \bibinfo{author}{\bibfnamefont{A.}~\bibnamefont{Lyapin}},
  \bibinfo{author}{\bibfnamefont{O.}~\bibnamefont{Stalgorova}},
  \bibinfo{author}{\bibfnamefont{E.}~\bibnamefont{Gromnitskaya}},
  \bibinfo{author}{\bibfnamefont{S.}~\bibnamefont{Popova}}, \bibnamefont{and}
  \bibinfo{author}{\bibfnamefont{O.}~\bibnamefont{Tsiok}},
  \bibinfo{journal}{Jour. Non-cryst. Solids} \textbf{\bibinfo{volume}{212}},
  \bibinfo{pages}{49} (\bibinfo{year}{1997}).

\bibitem[{\citenamefont{Mishima et~al.}(1985)\citenamefont{Mishima, Calvert,
  and Whalley}}]{Mishima-2}
\bibinfo{author}{\bibfnamefont{O.}~\bibnamefont{Mishima}},
  \bibinfo{author}{\bibfnamefont{L.~D.} \bibnamefont{Calvert}},
  \bibnamefont{and} \bibinfo{author}{\bibfnamefont{E.}~\bibnamefont{Whalley}},
  \bibinfo{journal}{Nature} \textbf{\bibinfo{volume}{314}}, \bibinfo{pages}{76
  } (\bibinfo{year}{1985}).

\bibitem[{\citenamefont{Daisenberger et~al.}(2007)\citenamefont{Daisenberger,
  Wilson, McMillan, Cabrera, Wilding, and Machon}}]{Daisenberger-Si}
\bibinfo{author}{\bibfnamefont{D.}~\bibnamefont{Daisenberger}},
  \bibinfo{author}{\bibfnamefont{M.}~\bibnamefont{Wilson}},
  \bibinfo{author}{\bibfnamefont{P.~F.} \bibnamefont{McMillan}},
  \bibinfo{author}{\bibfnamefont{R.~Q.} \bibnamefont{Cabrera}},
  \bibinfo{author}{\bibfnamefont{M.~C.} \bibnamefont{Wilding}},
  \bibnamefont{and} \bibinfo{author}{\bibfnamefont{D.}~\bibnamefont{Machon}},
  \bibinfo{journal}{Phys. Rev. B} \textbf{\bibinfo{volume}{75}},
  \bibinfo{pages}{224118} (\bibinfo{year}{2007}).

\bibitem[{\citenamefont{Principi et~al.}(2004)\citenamefont{Principi, DiCicco,
  Decremps, Polian, DePanfilis, and Filipponi}}]{Principi-Ge}
\bibinfo{author}{\bibfnamefont{E.}~\bibnamefont{Principi}},
  \bibinfo{author}{\bibfnamefont{A.}~\bibnamefont{DiCicco}},
  \bibinfo{author}{\bibfnamefont{F.}~\bibnamefont{Decremps}},
  \bibinfo{author}{\bibfnamefont{A.}~\bibnamefont{Polian}},
  \bibinfo{author}{\bibfnamefont{S.}~\bibnamefont{DePanfilis}},
  \bibnamefont{and}
  \bibinfo{author}{\bibfnamefont{A.}~\bibnamefont{Filipponi}},
  \bibinfo{journal}{Phys. Rev. B} \textbf{\bibinfo{volume}{69}},
  \bibinfo{pages}{201201} (\bibinfo{year}{2004}).

\bibitem[{\citenamefont{Huang and Kieffer}(2004)}]{Huang-Kieffer-SiO2-1}
\bibinfo{author}{\bibfnamefont{L.}~\bibnamefont{Huang}} \bibnamefont{and}
  \bibinfo{author}{\bibfnamefont{J.}~\bibnamefont{Kieffer}},
  \bibinfo{journal}{Phys. Rev. B} \textbf{\bibinfo{volume}{69}},
  \bibinfo{pages}{224203} (\bibinfo{year}{2004}).

\bibitem[{\citenamefont{Trachenko and Dove}(2003)}]{Trachenko-Dove}
\bibinfo{author}{\bibfnamefont{K.}~\bibnamefont{Trachenko}} \bibnamefont{and}
  \bibinfo{author}{\bibfnamefont{M.~T.} \bibnamefont{Dove}},
  \bibinfo{journal}{Phys. Rev. B} \textbf{\bibinfo{volume}{67}},
  \bibinfo{pages}{212203} (\bibinfo{year}{2003}).

\bibitem[{\citenamefont{Durandurdu and Drabold}(2002)}]{Durandurdu-GeSe2}
\bibinfo{author}{\bibfnamefont{M.}~\bibnamefont{Durandurdu}} \bibnamefont{and}
  \bibinfo{author}{\bibfnamefont{D.~A.} \bibnamefont{Drabold}},
  \bibinfo{journal}{Phys. Rev. B} \textbf{\bibinfo{volume}{65}},
  \bibinfo{pages}{104208} (\bibinfo{year}{2002}).

\bibitem[{\citenamefont{Trachenko et~al.}(2004)\citenamefont{Trachenko, Dove,
  Brazhkin, and El'kin}}]{Trachenko}
\bibinfo{author}{\bibfnamefont{K.}~\bibnamefont{Trachenko}},
  \bibinfo{author}{\bibfnamefont{M.~T.} \bibnamefont{Dove}},
  \bibinfo{author}{\bibfnamefont{V.}~\bibnamefont{Brazhkin}}, \bibnamefont{and}
  \bibinfo{author}{\bibfnamefont{F.~S.} \bibnamefont{El'kin}},
  \bibinfo{journal}{Phys. Rev. Lett.} \textbf{\bibinfo{volume}{93}},
  \bibinfo{pages}{135502} (\bibinfo{year}{2004}).

\bibitem[{\citenamefont{Sanloup}(2010)}]{HPC-Sanloup}
\bibinfo{author}{\bibfnamefont{C.}~\bibnamefont{Sanloup}}, in
  \emph{\bibinfo{booktitle}{High-Pressure Crystallography: From Fundamental
  Phenomena to Technological Applications}}, edited by
  \bibinfo{editor}{\bibfnamefont{E.}~\bibnamefont{Boldyreva}} \bibnamefont{and}
  \bibinfo{editor}{\bibfnamefont{P.}~\bibnamefont{Dera}}
  (\bibinfo{publisher}{Springer}, \bibinfo{address}{Dordrecht, The
  Netherlands}, \bibinfo{year}{2010}), chap.~\bibinfo{chapter}{37}, pp.
  \bibinfo{pages}{459--468}, ISBN \bibinfo{isbn}{978-90-481-9260-1(HB)}.

\bibitem[{\citenamefont{McMillan}(2004)}]{McMillan-AAT-review-1}
\bibinfo{author}{\bibfnamefont{P.~F.} \bibnamefont{McMillan}},
  \bibinfo{journal}{Jour. Mat. Chem.} \textbf{\bibinfo{volume}{14}},
  \bibinfo{pages}{1506} (\bibinfo{year}{2004}).

\bibitem[{\citenamefont{McMillan}(2002)}]{McMillan-AAT-review-2}
\bibinfo{author}{\bibfnamefont{P.~F.} \bibnamefont{McMillan}}, in
  \emph{\bibinfo{booktitle}{High Pressure Phenomena}}, edited by
  \bibinfo{editor}{\bibfnamefont{R.~J.} \bibnamefont{Hemley}} \bibnamefont{and}
  \bibinfo{editor}{\bibfnamefont{G.~L.} \bibnamefont{Chiarotti}}
  (\bibinfo{publisher}{IOS PRESS}, \bibinfo{address}{Amsterdam, The
  Netherlands}, \bibinfo{year}{2002}), pp. \bibinfo{pages}{511--541}, ISBN
  \bibinfo{isbn}{1-58603-269-0}.

\bibitem[{\citenamefont{Luo and Ruoff}(1993)}]{Luo-Ruoff}
\bibinfo{author}{\bibfnamefont{H.}~\bibnamefont{Luo}} \bibnamefont{and}
  \bibinfo{author}{\bibfnamefont{A.~L.} \bibnamefont{Ruoff}},
  \bibinfo{journal}{Phys. Rev. B} \textbf{\bibinfo{volume}{48}},
  \bibinfo{pages}{569} (\bibinfo{year}{1993}).

\bibitem[{\citenamefont{Sanloup et~al.}(2008)\citenamefont{Sanloup, Gregoryanz,
  Degtyareva, and Hanfland}}]{Sanloup}
\bibinfo{author}{\bibfnamefont{C.}~\bibnamefont{Sanloup}},
  \bibinfo{author}{\bibfnamefont{E.}~\bibnamefont{Gregoryanz}},
  \bibinfo{author}{\bibfnamefont{O.}~\bibnamefont{Degtyareva}},
  \bibnamefont{and} \bibinfo{author}{\bibfnamefont{M.}~\bibnamefont{Hanfland}},
  \bibinfo{journal}{Phys. Rev. Lett.} \textbf{\bibinfo{volume}{100}},
  \bibinfo{pages}{075701} (\bibinfo{year}{2008}).

\bibitem[{\citenamefont{Gregoryanz et~al.}()\citenamefont{Gregoryanz, Sanloup,
  Degtyareva, Hemley, and Hanfland}}]{Gregoryanz-unpublished}
\bibinfo{author}{\bibfnamefont{E.}~\bibnamefont{Gregoryanz}},
  \bibinfo{author}{\bibfnamefont{C.}~\bibnamefont{Sanloup}},
  \bibinfo{author}{\bibfnamefont{O.}~\bibnamefont{Degtyareva}},
  \bibinfo{author}{\bibfnamefont{R.~J.} \bibnamefont{Hemley}},
  \bibnamefont{and} \bibinfo{author}{\bibfnamefont{M.}~\bibnamefont{Hanfland}},
  \bibinfo{howpublished}{personal communication}.

\bibitem[{\citenamefont{Meyer}(1976)}]{Meyer}
\bibinfo{author}{\bibfnamefont{B.}~\bibnamefont{Meyer}},
  \bibinfo{journal}{Chemical Reviews} \textbf{\bibinfo{volume}{76}},
  \bibinfo{pages}{367} (\bibinfo{year}{1976}).

\bibitem[{\citenamefont{Degtyareva
  et~al.}(2005{\natexlab{a}})\citenamefont{Degtyareva, Gregoryanz, Somayazulu,
  Dera, Mao, and Hemley}}]{Deg-1}
\bibinfo{author}{\bibfnamefont{O.}~\bibnamefont{Degtyareva}},
  \bibinfo{author}{\bibfnamefont{E.}~\bibnamefont{Gregoryanz}},
  \bibinfo{author}{\bibfnamefont{M.}~\bibnamefont{Somayazulu}},
  \bibinfo{author}{\bibfnamefont{P.}~\bibnamefont{Dera}},
  \bibinfo{author}{\bibfnamefont{H.~K.} \bibnamefont{Mao}}, \bibnamefont{and}
  \bibinfo{author}{\bibfnamefont{R.}~\bibnamefont{Hemley}},
  \bibinfo{journal}{Nature materials} \textbf{\bibinfo{volume}{4}},
  \bibinfo{pages}{152} (\bibinfo{year}{2005}{\natexlab{a}}).

\bibitem[{\citenamefont{Degtyareva
  et~al.}(2007{\natexlab{a}})\citenamefont{Degtyareva, Hern\'{a}ndez, Serrano,
  Somayazulu, Mao, Gregoryanz, and Hemley}}]{Deg-2}
\bibinfo{author}{\bibfnamefont{O.}~\bibnamefont{Degtyareva}},
  \bibinfo{author}{\bibfnamefont{E.}~\bibnamefont{Hern\'{a}ndez}},
  \bibinfo{author}{\bibfnamefont{J.}~\bibnamefont{Serrano}},
  \bibinfo{author}{\bibfnamefont{M.}~\bibnamefont{Somayazulu}},
  \bibinfo{author}{\bibfnamefont{H.~K.} \bibnamefont{Mao}},
  \bibinfo{author}{\bibfnamefont{E.}~\bibnamefont{Gregoryanz}},
  \bibnamefont{and} \bibinfo{author}{\bibfnamefont{R.}~\bibnamefont{Hemley}},
  \bibinfo{journal}{J. Chem. Phys.} \textbf{\bibinfo{volume}{126}},
  \bibinfo{pages}{084503} (\bibinfo{year}{2007}{\natexlab{a}}).

\bibitem[{\citenamefont{Degtyareva
  et~al.}(2005{\natexlab{b}})\citenamefont{Degtyareva, Gregoryanz, Somayazulu,
  Mao, and Hemley}}]{Deg-3}
\bibinfo{author}{\bibfnamefont{O.}~\bibnamefont{Degtyareva}},
  \bibinfo{author}{\bibfnamefont{E.}~\bibnamefont{Gregoryanz}},
  \bibinfo{author}{\bibfnamefont{M.}~\bibnamefont{Somayazulu}},
  \bibinfo{author}{\bibfnamefont{H.~K.} \bibnamefont{Mao}}, \bibnamefont{and}
  \bibinfo{author}{\bibfnamefont{R.~J.} \bibnamefont{Hemley}},
  \bibinfo{journal}{Phys. Rev. B} \textbf{\bibinfo{volume}{71}},
  \bibinfo{pages}{214104} (\bibinfo{year}{2005}{\natexlab{b}}).

\bibitem[{\citenamefont{Degtyareva
  et~al.}(2007{\natexlab{b}})\citenamefont{Degtyareva, Magnitskaya, Kohanoff,
  Profeta, Scandolo, Hanfland, McMahon, and Gregoryanz}}]{Deg-4}
\bibinfo{author}{\bibfnamefont{O.}~\bibnamefont{Degtyareva}},
  \bibinfo{author}{\bibfnamefont{M.~V.} \bibnamefont{Magnitskaya}},
  \bibinfo{author}{\bibfnamefont{J.}~\bibnamefont{Kohanoff}},
  \bibinfo{author}{\bibfnamefont{G.}~\bibnamefont{Profeta}},
  \bibinfo{author}{\bibfnamefont{S.}~\bibnamefont{Scandolo}},
  \bibinfo{author}{\bibfnamefont{M.}~\bibnamefont{Hanfland}},
  \bibinfo{author}{\bibfnamefont{M.~I.} \bibnamefont{McMahon}},
  \bibnamefont{and}
  \bibinfo{author}{\bibfnamefont{E.}~\bibnamefont{Gregoryanz}},
  \bibinfo{journal}{Phys. Rev. Lett.} \textbf{\bibinfo{volume}{99}},
  \bibinfo{pages}{155505} (\bibinfo{year}{2007}{\natexlab{b}}).

\bibitem[{\citenamefont{Degtyareva
  et~al.}(2005{\natexlab{c}})\citenamefont{Degtyareva, Gregoryanz, Mao, and
  Hemley}}]{Deg-5}
\bibinfo{author}{\bibfnamefont{O.}~\bibnamefont{Degtyareva}},
  \bibinfo{author}{\bibfnamefont{E.}~\bibnamefont{Gregoryanz}},
  \bibinfo{author}{\bibfnamefont{H.~K.} \bibnamefont{Mao}}, \bibnamefont{and}
  \bibinfo{author}{\bibfnamefont{R.~J.} \bibnamefont{Hemley}},
  \bibinfo{journal}{High Press. Res.} \textbf{\bibinfo{volume}{25}},
  \bibinfo{pages}{17} (\bibinfo{year}{2005}{\natexlab{c}}).

\bibitem[{\citenamefont{Hejny et~al.}(2005)\citenamefont{Hejny, Lundegaard,
  Falconi, McMahon, and Hanfland}}]{Hejny}
\bibinfo{author}{\bibfnamefont{C.}~\bibnamefont{Hejny}},
  \bibinfo{author}{\bibfnamefont{L.~F.} \bibnamefont{Lundegaard}},
  \bibinfo{author}{\bibfnamefont{S.}~\bibnamefont{Falconi}},
  \bibinfo{author}{\bibfnamefont{M.~I.} \bibnamefont{McMahon}},
  \bibnamefont{and} \bibinfo{author}{\bibfnamefont{M.}~\bibnamefont{Hanfland}},
  \bibinfo{journal}{Phys. Rev. B} \textbf{\bibinfo{volume}{71}},
  \bibinfo{pages}{020101} (\bibinfo{year}{2005}).

\bibitem[{\citenamefont{Crapanzano et~al.}(2005)\citenamefont{Crapanzano,
  Crichton, Monaco, Bellissent, and Mezouar}}]{Crapanzano}
\bibinfo{author}{\bibfnamefont{L.}~\bibnamefont{Crapanzano}},
  \bibinfo{author}{\bibfnamefont{W.~A.} \bibnamefont{Crichton}},
  \bibinfo{author}{\bibfnamefont{G.}~\bibnamefont{Monaco}},
  \bibinfo{author}{\bibfnamefont{R.}~\bibnamefont{Bellissent}},
  \bibnamefont{and} \bibinfo{author}{\bibfnamefont{M.}~\bibnamefont{Mezouar}},
  \bibinfo{journal}{Nature materials} \textbf{\bibinfo{volume}{4}},
  \bibinfo{pages}{550} (\bibinfo{year}{2005}).

\bibitem[{\citenamefont{Zakharov and Cohen}(1995)}]{Zakharov-Cohen}
\bibinfo{author}{\bibfnamefont{O.}~\bibnamefont{Zakharov}} \bibnamefont{and}
  \bibinfo{author}{\bibfnamefont{M.~L.} \bibnamefont{Cohen}},
  \bibinfo{journal}{Phys. Rev. B} \textbf{\bibinfo{volume}{52}},
  \bibinfo{pages}{12572} (\bibinfo{year}{1995}).

\bibitem[{\citenamefont{Oganov and Glass}(2006)}]{Oganov}
\bibinfo{author}{\bibfnamefont{A.~R.} \bibnamefont{Oganov}} \bibnamefont{and}
  \bibinfo{author}{\bibfnamefont{C.~W.} \bibnamefont{Glass}},
  \bibinfo{journal}{J. Chem. Phys.} \textbf{\bibinfo{volume}{124}},
  \bibinfo{pages}{244704} (\bibinfo{year}{2006}).

\bibitem[{\citenamefont{Pastorino and Gamba}(2003)}]{Pastorino-Gamba-1}
\bibinfo{author}{\bibfnamefont{C.}~\bibnamefont{Pastorino}} \bibnamefont{and}
  \bibinfo{author}{\bibfnamefont{Z.}~\bibnamefont{Gamba}}, \bibinfo{journal}{J.
  Chem. Phys.} \textbf{\bibinfo{volume}{119}}, \bibinfo{pages}{2147}
  (\bibinfo{year}{2003}).

\bibitem[{\citenamefont{Rudin and Liu}(1999)}]{Rudin-Liu}
\bibinfo{author}{\bibfnamefont{S.~P.} \bibnamefont{Rudin}} \bibnamefont{and}
  \bibinfo{author}{\bibfnamefont{A.~Y.} \bibnamefont{Liu}},
  \bibinfo{journal}{Phys. Rev. Lett.} \textbf{\bibinfo{volume}{83}},
  \bibinfo{pages}{3049} (\bibinfo{year}{1999}).

\bibitem[{\citenamefont{Fujihisa et~al.}(2004)\citenamefont{Fujihisa, Akahama,
  Kawamura, Yamawaki, Sakashita, Yamada, Honda, and LeBihan}}]{Fujihisa}
\bibinfo{author}{\bibfnamefont{H.}~\bibnamefont{Fujihisa}},
  \bibinfo{author}{\bibfnamefont{Y.}~\bibnamefont{Akahama}},
  \bibinfo{author}{\bibfnamefont{H.}~\bibnamefont{Kawamura}},
  \bibinfo{author}{\bibfnamefont{H.}~\bibnamefont{Yamawaki}},
  \bibinfo{author}{\bibfnamefont{M.}~\bibnamefont{Sakashita}},
  \bibinfo{author}{\bibfnamefont{T.}~\bibnamefont{Yamada}},
  \bibinfo{author}{\bibfnamefont{K.}~\bibnamefont{Honda}}, \bibnamefont{and}
  \bibinfo{author}{\bibfnamefont{T.}~\bibnamefont{LeBihan}},
  \bibinfo{journal}{Phys. Rev. B} \textbf{\bibinfo{volume}{70}},
  \bibinfo{pages}{134106} (\bibinfo{year}{2004}).

\bibitem[{\citenamefont{Nishikawa}(2008)}]{Nishikawa}
\bibinfo{author}{\bibfnamefont{A.}~\bibnamefont{Nishikawa}},
  \bibinfo{journal}{Journal of Physics: Conference Series}
  \textbf{\bibinfo{volume}{121}}, \bibinfo{pages}{012008}
  (\bibinfo{year}{2008}).

\bibitem[{\citenamefont{Luo et~al.}(1993)\citenamefont{Luo, Greene, and
  Ruoff}}]{Luo-Greene-Ruoff}
\bibinfo{author}{\bibfnamefont{H.}~\bibnamefont{Luo}},
  \bibinfo{author}{\bibfnamefont{R.~G.} \bibnamefont{Greene}},
  \bibnamefont{and} \bibinfo{author}{\bibfnamefont{A.~L.} \bibnamefont{Ruoff}},
  \bibinfo{journal}{Phys. Rev. Lett.} \textbf{\bibinfo{volume}{71}},
  \bibinfo{pages}{2943} (\bibinfo{year}{1993}).

\bibitem[{\citenamefont{Nishikawa et~al.}(1999)\citenamefont{Nishikawa,
  Niizeki, and Shindo}}]{Nishikawa-Niizeki-Shindo}
\bibinfo{author}{\bibfnamefont{A.}~\bibnamefont{Nishikawa}},
  \bibinfo{author}{\bibfnamefont{K.}~\bibnamefont{Niizeki}}, \bibnamefont{and}
  \bibinfo{author}{\bibfnamefont{K.}~\bibnamefont{Shindo}},
  \bibinfo{journal}{Phys. stat. sol. (b)} \textbf{\bibinfo{volume}{211}},
  \bibinfo{pages}{373} (\bibinfo{year}{1999}).

\bibitem[{\citenamefont{Luo et~al.}(1991)\citenamefont{Luo, Desgreniers, Vohra,
  and Ruoff}}]{Luo-Vohra}
\bibinfo{author}{\bibfnamefont{H.}~\bibnamefont{Luo}},
  \bibinfo{author}{\bibfnamefont{S.}~\bibnamefont{Desgreniers}},
  \bibinfo{author}{\bibfnamefont{Y.~K.} \bibnamefont{Vohra}}, \bibnamefont{and}
  \bibinfo{author}{\bibfnamefont{A.~L.} \bibnamefont{Ruoff}},
  \bibinfo{journal}{Phys. Rev. Lett.} \textbf{\bibinfo{volume}{67}},
  \bibinfo{pages}{2998} (\bibinfo{year}{1991}).

\bibitem[{\citenamefont{Dunn and Bundy}(1977)}]{Dunn-Bundy}
\bibinfo{author}{\bibfnamefont{K.~J.} \bibnamefont{Dunn}} \bibnamefont{and}
  \bibinfo{author}{\bibfnamefont{F.~P.} \bibnamefont{Bundy}},
  \bibinfo{journal}{J. Chem. Phys.} \textbf{\bibinfo{volume}{67}},
  \bibinfo{pages}{5048} (\bibinfo{year}{1977}).

\bibitem[{\citenamefont{Gregoryanz et~al.}(2002)\citenamefont{Gregoryanz,
  Struzhkin, Hemley, Eremets, Mao, and Timofeev}}]{Gregoryanz-supercond}
\bibinfo{author}{\bibfnamefont{E.}~\bibnamefont{Gregoryanz}},
  \bibinfo{author}{\bibfnamefont{V.~V.} \bibnamefont{Struzhkin}},
  \bibinfo{author}{\bibfnamefont{R.~J.} \bibnamefont{Hemley}},
  \bibinfo{author}{\bibfnamefont{M.~I.} \bibnamefont{Eremets}},
  \bibinfo{author}{\bibfnamefont{H.~K.} \bibnamefont{Mao}}, \bibnamefont{and}
  \bibinfo{author}{\bibfnamefont{Y.~A.} \bibnamefont{Timofeev}},
  \bibinfo{journal}{Phys. Rev. B} \textbf{\bibinfo{volume}{65}},
  \bibinfo{pages}{064504} (\bibinfo{year}{2002}).

\bibitem[{\citenamefont{Bellissent et~al.}(1990)\citenamefont{Bellissent,
  Descotes, Bou\'{e}, and Pfeuty}}]{Bellissent}
\bibinfo{author}{\bibfnamefont{R.}~\bibnamefont{Bellissent}},
  \bibinfo{author}{\bibfnamefont{L.}~\bibnamefont{Descotes}},
  \bibinfo{author}{\bibfnamefont{F.}~\bibnamefont{Bou\'{e}}}, \bibnamefont{and}
  \bibinfo{author}{\bibfnamefont{P.}~\bibnamefont{Pfeuty}},
  \bibinfo{journal}{Phys. Rev. B} \textbf{\bibinfo{volume}{41}},
  \bibinfo{pages}{2135} (\bibinfo{year}{1990}).

\bibitem[{\citenamefont{Kalampounias et~al.}(2003)\citenamefont{Kalampounias,
  Andrikopoulos, and Yannopoulos}}]{Kalampounias}
\bibinfo{author}{\bibfnamefont{A.~G.} \bibnamefont{Kalampounias}},
  \bibinfo{author}{\bibfnamefont{K.~S.} \bibnamefont{Andrikopoulos}},
  \bibnamefont{and} \bibinfo{author}{\bibfnamefont{S.~N.}
  \bibnamefont{Yannopoulos}}, \bibinfo{journal}{J. Chem. Phys.}
  \textbf{\bibinfo{volume}{118}}, \bibinfo{pages}{8460} (\bibinfo{year}{2003}).

\bibitem[{\citenamefont{Scopigno et~al.}(2007)\citenamefont{Scopigno,
  Yannopoulos, Scarponi, Andrikopoulos, Fioretto, and Ruocco}}]{Scopigno}
\bibinfo{author}{\bibfnamefont{T.}~\bibnamefont{Scopigno}},
  \bibinfo{author}{\bibfnamefont{S.~N.} \bibnamefont{Yannopoulos}},
  \bibinfo{author}{\bibfnamefont{F.}~\bibnamefont{Scarponi}},
  \bibinfo{author}{\bibfnamefont{K.~S.} \bibnamefont{Andrikopoulos}},
  \bibinfo{author}{\bibfnamefont{D.}~\bibnamefont{Fioretto}}, \bibnamefont{and}
  \bibinfo{author}{\bibfnamefont{G.}~\bibnamefont{Ruocco}},
  \bibinfo{journal}{Phys. Rev. Lett.} \textbf{\bibinfo{volume}{99}},
  \bibinfo{pages}{025701} (\bibinfo{year}{2007}).

\bibitem[{\citenamefont{Monaco et~al.}(2005)\citenamefont{Monaco, Crapanzano,
  Bellissent, Crichton, Fioretto, Mezouar, Scarponi, and Verbeni}}]{Monaco}
\bibinfo{author}{\bibfnamefont{G.}~\bibnamefont{Monaco}},
  \bibinfo{author}{\bibfnamefont{L.}~\bibnamefont{Crapanzano}},
  \bibinfo{author}{\bibfnamefont{R.}~\bibnamefont{Bellissent}},
  \bibinfo{author}{\bibfnamefont{W.}~\bibnamefont{Crichton}},
  \bibinfo{author}{\bibfnamefont{D.}~\bibnamefont{Fioretto}},
  \bibinfo{author}{\bibfnamefont{M.}~\bibnamefont{Mezouar}},
  \bibinfo{author}{\bibfnamefont{F.}~\bibnamefont{Scarponi}}, \bibnamefont{and}
  \bibinfo{author}{\bibfnamefont{R.}~\bibnamefont{Verbeni}},
  \bibinfo{journal}{Phys. Rev. Lett.} \textbf{\bibinfo{volume}{95}},
  \bibinfo{pages}{255502} (\bibinfo{year}{2005}).

\bibitem[{\citenamefont{Springborg and Jones}(1986)}]{Springborg}
\bibinfo{author}{\bibfnamefont{M.}~\bibnamefont{Springborg}} \bibnamefont{and}
  \bibinfo{author}{\bibfnamefont{R.~O.} \bibnamefont{Jones}},
  \bibinfo{journal}{Phys. Rev. Lett.} \textbf{\bibinfo{volume}{57}},
  \bibinfo{pages}{1145} (\bibinfo{year}{1986}).

\bibitem[{\citenamefont{Tse and Klug}(1999)}]{Tse-Klug-S}
\bibinfo{author}{\bibfnamefont{J.~S.} \bibnamefont{Tse}} \bibnamefont{and}
  \bibinfo{author}{\bibfnamefont{D.~D.} \bibnamefont{Klug}},
  \bibinfo{journal}{Phys. Rev. B} \textbf{\bibinfo{volume}{59}},
  \bibinfo{pages}{34} (\bibinfo{year}{1999}).

\bibitem[{\citenamefont{Jones and Ballone}(2004)}]{Jones-Ballone-2}
\bibinfo{author}{\bibfnamefont{R.~O.} \bibnamefont{Jones}} \bibnamefont{and}
  \bibinfo{author}{\bibfnamefont{P.}~\bibnamefont{Ballone}}, in
  \emph{\bibinfo{booktitle}{3rd International Conference: Computational
  Modeling and Simulations of materials}}, edited by
  \bibinfo{editor}{\bibfnamefont{P.}~\bibnamefont{Vincenzini}}
  \bibnamefont{and} \bibinfo{editor}{\bibfnamefont{A.}~\bibnamefont{Lami}}
  (\bibinfo{year}{2004}), pp. \bibinfo{pages}{281--288}.

\bibitem[{\citenamefont{Yu et~al.}(2009)\citenamefont{Yu, Wang, Wang, Lin, Liu,
  Hong, and Bai}}]{Yu-S}
\bibinfo{author}{\bibfnamefont{P.}~\bibnamefont{Yu}},
  \bibinfo{author}{\bibfnamefont{W.~H.} \bibnamefont{Wang}},
  \bibinfo{author}{\bibfnamefont{R.~J.} \bibnamefont{Wang}},
  \bibinfo{author}{\bibfnamefont{S.~X.} \bibnamefont{Lin}},
  \bibinfo{author}{\bibfnamefont{X.~R.} \bibnamefont{Liu}},
  \bibinfo{author}{\bibfnamefont{S.~M.} \bibnamefont{Hong}}, \bibnamefont{and}
  \bibinfo{author}{\bibfnamefont{H.~Y.} \bibnamefont{Bai}},
  \bibinfo{journal}{Appl. Phys. Lett.} \textbf{\bibinfo{volume}{94}},
  \bibinfo{pages}{011910} (\bibinfo{year}{2009}).

\bibitem[{\citenamefont{Akahama et~al.}(1993)\citenamefont{Akahama, Kobayashi,
  and Kawamura}}]{Akahama}
\bibinfo{author}{\bibfnamefont{Y.}~\bibnamefont{Akahama}},
  \bibinfo{author}{\bibfnamefont{M.}~\bibnamefont{Kobayashi}},
  \bibnamefont{and} \bibinfo{author}{\bibfnamefont{H.}~\bibnamefont{Kawamura}},
  \bibinfo{journal}{Phys. Rev. B} \textbf{\bibinfo{volume}{48}},
  \bibinfo{pages}{6862} (\bibinfo{year}{1993}).

\bibitem[{\citenamefont{Kresse and
  Furthm\"{u}ller}(1996{\natexlab{a}})}]{VASP-1}
\bibinfo{author}{\bibfnamefont{G.}~\bibnamefont{Kresse}} \bibnamefont{and}
  \bibinfo{author}{\bibfnamefont{J.}~\bibnamefont{Furthm\"{u}ller}},
  \bibinfo{journal}{Phys. Rev. B} \textbf{\bibinfo{volume}{54}},
  \bibinfo{pages}{11169} (\bibinfo{year}{1996}{\natexlab{a}}).

\bibitem[{\citenamefont{Kresse and
  Furthm\"{u}ller}(1996{\natexlab{b}})}]{VASP-2}
\bibinfo{author}{\bibfnamefont{G.}~\bibnamefont{Kresse}} \bibnamefont{and}
  \bibinfo{author}{\bibfnamefont{J.}~\bibnamefont{Furthm\"{u}ller}},
  \bibinfo{journal}{Comput. Mat. Sci.} \textbf{\bibinfo{volume}{6}},
  \bibinfo{pages}{15} (\bibinfo{year}{1996}{\natexlab{b}}).

\bibitem[{\citenamefont{Kresse and Hafner}(1993)}]{VASP-3}
\bibinfo{author}{\bibfnamefont{G.}~\bibnamefont{Kresse}} \bibnamefont{and}
  \bibinfo{author}{\bibfnamefont{J.}~\bibnamefont{Hafner}},
  \bibinfo{journal}{Phys. Rev. B} \textbf{\bibinfo{volume}{47}},
  \bibinfo{pages}{558} (\bibinfo{year}{1993}).

\bibitem[{\citenamefont{Kresse and Hafner}(1994)}]{VASP-4}
\bibinfo{author}{\bibfnamefont{G.}~\bibnamefont{Kresse}} \bibnamefont{and}
  \bibinfo{author}{\bibfnamefont{J.}~\bibnamefont{Hafner}},
  \bibinfo{journal}{Phys. Rev. B} \textbf{\bibinfo{volume}{49}},
  \bibinfo{pages}{14251} (\bibinfo{year}{1994}).

\bibitem[{\citenamefont{Berendsen et~al.}(1984)\citenamefont{Berendsen, Postma,
  van Gunsteren, DiNola, and Haak}}]{Berendsen}
\bibinfo{author}{\bibfnamefont{H.~J.~C.} \bibnamefont{Berendsen}},
  \bibinfo{author}{\bibfnamefont{J.~P.~M.} \bibnamefont{Postma}},
  \bibinfo{author}{\bibfnamefont{W.~F.} \bibnamefont{van Gunsteren}},
  \bibinfo{author}{\bibfnamefont{A.}~\bibnamefont{DiNola}}, \bibnamefont{and}
  \bibinfo{author}{\bibfnamefont{J.~R.} \bibnamefont{Haak}},
  \bibinfo{journal}{J. Chem. Phys.} \textbf{\bibinfo{volume}{81}},
  \bibinfo{pages}{3684} (\bibinfo{year}{1984}).

\bibitem[{\citenamefont{Bl\"{o}chl}(1994)}]{PAW}
\bibinfo{author}{\bibfnamefont{P.~E.} \bibnamefont{Bl\"{o}chl}},
  \bibinfo{journal}{Phys. Rev. B} \textbf{\bibinfo{volume}{50}},
  \bibinfo{pages}{17953 } (\bibinfo{year}{1994}).

\bibitem[{\citenamefont{Kresse and Joubert}(1999)}]{VASP-PAW}
\bibinfo{author}{\bibfnamefont{G.}~\bibnamefont{Kresse}} \bibnamefont{and}
  \bibinfo{author}{\bibfnamefont{D.}~\bibnamefont{Joubert}},
  \bibinfo{journal}{Phys. Rev. B} \textbf{\bibinfo{volume}{59}},
  \bibinfo{pages}{1758 } (\bibinfo{year}{1999}).

\bibitem[{\citenamefont{Perdew et~al.}(1996)\citenamefont{Perdew, Burke, and
  Ernzerhof}}]{PBE}
\bibinfo{author}{\bibfnamefont{J.~P.} \bibnamefont{Perdew}},
  \bibinfo{author}{\bibfnamefont{K.}~\bibnamefont{Burke}}, \bibnamefont{and}
  \bibinfo{author}{\bibfnamefont{M.}~\bibnamefont{Ernzerhof}},
  \bibinfo{journal}{Phys. Rev. Lett.} \textbf{\bibinfo{volume}{77}},
  \bibinfo{pages}{3865} (\bibinfo{year}{1996}).

\bibitem[{\citenamefont{Kohn and Sham}(1965)}]{KS}
\bibinfo{author}{\bibfnamefont{W.}~\bibnamefont{Kohn}} \bibnamefont{and}
  \bibinfo{author}{\bibfnamefont{L.~J.} \bibnamefont{Sham}},
  \bibinfo{journal}{Phys. Rev.} \textbf{\bibinfo{volume}{140}},
  \bibinfo{pages}{1133} (\bibinfo{year}{1965}).

\bibitem[{\citenamefont{Jones and Ballone}(2003)}]{Jones-Ballone-1}
\bibinfo{author}{\bibfnamefont{R.~O.} \bibnamefont{Jones}} \bibnamefont{and}
  \bibinfo{author}{\bibfnamefont{P.}~\bibnamefont{Ballone}},
  \bibinfo{journal}{J. Chem. Phys.} \textbf{\bibinfo{volume}{118}},
  \bibinfo{pages}{9257} (\bibinfo{year}{2003}).

\bibitem[{\citenamefont{Hohl et~al.}(1988)\citenamefont{Hohl, Jones, Car, and
  Parrinello}}]{Hohl-Jones}
\bibinfo{author}{\bibfnamefont{D.}~\bibnamefont{Hohl}},
  \bibinfo{author}{\bibfnamefont{R.~O.} \bibnamefont{Jones}},
  \bibinfo{author}{\bibfnamefont{R.}~\bibnamefont{Car}}, \bibnamefont{and}
  \bibinfo{author}{\bibfnamefont{M.}~\bibnamefont{Parrinello}},
  \bibinfo{journal}{J. Chem. Phys.} \textbf{\bibinfo{volume}{89}},
  \bibinfo{pages}{6823} (\bibinfo{year}{1988}).

\bibitem[{Jmo()}]{Jmol}
\emph{\bibinfo{title}{Jmol: an open-source java viewer for chemical structures
  in 3d. http://www.jmol.org/}}.

\bibitem[{\citenamefont{Goncharov et~al.}(2000)\citenamefont{Goncharov,
  Gregoryanz, Mao, Liu, and Hemley}}]{Goncharov-N}
\bibinfo{author}{\bibfnamefont{A.~F.} \bibnamefont{Goncharov}},
  \bibinfo{author}{\bibfnamefont{E.}~\bibnamefont{Gregoryanz}},
  \bibinfo{author}{\bibfnamefont{H.~K.} \bibnamefont{Mao}},
  \bibinfo{author}{\bibfnamefont{Z.}~\bibnamefont{Liu}}, \bibnamefont{and}
  \bibinfo{author}{\bibfnamefont{R.~J.} \bibnamefont{Hemley}},
  \bibinfo{journal}{Phys. Rev. Lett.} \textbf{\bibinfo{volume}{85}},
  \bibinfo{pages}{1262} (\bibinfo{year}{2000}).

\bibitem[{\citenamefont{Santoro et~al.}(2006)\citenamefont{Santoro, Gorelli,
  Bini, Ruocco, Scandolo, and Crichton}}]{Santoro-1}
\bibinfo{author}{\bibfnamefont{M.}~\bibnamefont{Santoro}},
  \bibinfo{author}{\bibfnamefont{F.~A.} \bibnamefont{Gorelli}},
  \bibinfo{author}{\bibfnamefont{R.}~\bibnamefont{Bini}},
  \bibinfo{author}{\bibfnamefont{G.}~\bibnamefont{Ruocco}},
  \bibinfo{author}{\bibfnamefont{S.}~\bibnamefont{Scandolo}}, \bibnamefont{and}
  \bibinfo{author}{\bibfnamefont{W.~A.} \bibnamefont{Crichton}},
  \bibinfo{journal}{Nature} \textbf{\bibinfo{volume}{441}},
  \bibinfo{pages}{857} (\bibinfo{year}{2006}).

\end{thebibliography}

\end{document}